\documentclass[twocolumn,prb,showpacs,aps]{revtex4} 
\usepackage{graphicx}
\usepackage{amsmath}
\usepackage{amsbsy}
\usepackage{mathbbol}

\usepackage{color}
\graphicspath{{figures/}}

\begin{document}

\title{Spin accumulation in diffusive conductors \\
    with Rashba and Dresselhaus spin-orbit interaction}

\author{Mathias Duckheim}
\email{mathias.duckheim@unibas.ch}
\author{Daniel Loss}
\affiliation{Department of Physics, University of Basel, CH-4056
Basel, Switzerland}

\author{Matthias Scheid}
\email{matthias.scheid@uni-r.de}
\author{Klaus Richter}
\affiliation{Institut f\"ur Theoretische Physik, Universit\"at Regensburg, D-93040 Regensburg, Germany}
\author{\.{I}nan\c{c} Adagideli}
\affiliation{Faculty of Engineering and Natural Sciences, Sabanci University, 
Tuzla, Istanbul 34956, Turkey }
\author{Philippe Jacquod}
\affiliation{Department of Physics, University of Arizona, 1118 E. 4th Street, Tucson, AZ 85721, USA}

\date{\today}
%



\begin{abstract}
We calculate the electrically induced spin accumulation in diffusive
systems due to both
Rashba (with strength $\alpha)$ and 
Dresselhaus (with strength $\beta)$ spin-orbit interaction.
Using a diffusion equation approach we find that magnetoelectric effects disappear and that
there is thus no spin accumulation when both interactions have the same strength, $\alpha=\pm \beta$.
In thermodynamically large systems, the finite 
spin accumulation predicted by Chaplik, Entin and Magarill,
[Physica E {\bf 13}, 744 (2002)] and by Trushin and Schliemann [Phys. Rev. B {\bf 75}, 155323 (2007)]
is recovered an infinitesimally small distance away 
from the singular point $\alpha=\pm \beta$.
We show however that the singularity is broadened and that the suppression of
spin accumulation becomes physically relevant (i) in finite-sized systems of size $L$, (ii)
in the presence of a cubic Dresselhaus interaction of strength $\gamma$,
or (iii) for finite frequency measurements.  We obtain
the parametric range over which the magnetoelectric effect is suppressed in these three instances as
(i) $|\alpha|-|\beta| \lesssim 1/mL$, (ii)$|\alpha|-|\beta| \lesssim \gamma p_{\rm F}^2$, and
  (iii) $|\alpha|-|\beta| \lesssim  \sqrt{\omega/m p_{\rm F}\ell}$ with $\ell$ the elastic mean
  free path and $p_{\rm F}$ the Fermi momentum.  We attribute the absence
of spin accumulation close to $\alpha=\pm \beta$ to the underlying U (1) 
symmetry. We illustrate and confirm our predictions numerically.
 \end{abstract}

\maketitle

\section{Introduction}

Many recent works have explored mechanisms to generate spin
accumulations and spin currents by passing electric currents through
spin-orbit coupled electronic systems. On
the theoretical side, two related mechanisms have been
proposed in disordered metals as alternatives to spin
injection via ferromagnets or by optical means. They
are current-induced transverse spin currents or voltages,
a.k.a. the spin Hall effect,~\cite{Dya71a,Zha00,Sin04,Mur04,Ino04,Mis04,Bur04,Ada05,Nik05,Sch05,Rai05,Cha05}
and current-induced spin
polarization (CISP).~\cite{Vas79,Lev85,Aro89,Ede90,Cha02,Tru07} 
The interplay between the
two effects has been investigated in Ref.~\onlinecite{Ada07}. These effects
have been to some extent demonstrated experimentally,~\cite{Wun05,Kat04,Sai06,Zha06,Sek08}
and recent theoretical works have extended
them to include the mesoscopic regime, where fluctuations
of both longitudinal and transverse spin currents in
mesoscopic ballistic and diffusive systems are being investigated.~\cite{Bar07,Naz07,Kri08,Duc08,Dra08}
Most remarkably, it has been found that
the same universality that applies to mesoscopic charge
transport~\cite{als} also applies to mesoscopic spin transport.~\cite{Bar07}

The main focus of these theoretical as well as experimental efforts is
to use spin-orbit interactions (SOI) as a way to couple external electric
fields to electronic spins, the hope being to generate, manipulate
and/or measure spin currents and accumulations by purely electrical
means.  SOI, however, also has the undesired
effect of randomizing electronic spins.~\cite{Dya71b} This dichotomy
theoretically limits the use of SOI based magnetoelectric
devices as components of information processors to the regime
where the size of the device is much less than the spin
relaxation length. A way to increase the spin relaxation length
has been proposed in Ref.~\onlinecite{Sch03} for systems
which exhibit SOI of both the Rashba~\cite{Ras60}
\begin{equation}\label{eq:rashbaH}
H_{\rm R} = \alpha (p_x \sigma_y - p_y \sigma_x) \, ,
\end{equation}
and the Dresselhaus type~\cite{Dre55}
\begin{equation}\label{eq:dresselH}
H_{\rm D} = \beta (p_x \sigma_x - p_y \sigma_y) \, ,
\end{equation}
where $\sigma_{x,y}$ are Pauli matrices.
When the two interactions have equal strength, $\alpha = \pm \beta$, the
SOI rotates electron spins around a single, fixed axis. The spin along this 
axis
becomes conserved while spins along the perpendicular directions
undergo a deterministic rotation that depends only on the starting
and endpoints of their trajectory. In particular, spins are
not rotated along closed trajectories, therefore mesoscopic systems
exhibit negative magnetoresistance
when $\alpha=\pm \beta$,~\cite{Pik95,Zai05,Sch08} i.e. weak localization and not weak antilocalization, just as if SOI were absent. An effective spin randomization still occurs
if the system is connected to many external transport channels, where
uncertainties in the position of injection and exit translate into 
uncertainties in the spin rotation angle, unless injected electrons
are prepared as spin-eigenstates of the Hamiltonian.~\cite{Sch03}
In Ref.~\onlinecite{Sch03} (see, in particular, Eq. (7) therein)
spatially periodic modes in diffusive systems have been first described
for the case of equal strengths $\alpha = \pm \beta$, with spatial
period given by the spin orbit length. These modes are long-lived for
these particular SOI strengths (and in the absence of cubic SOI) and are
thus referred to as persistent spin helix,~\cite{Ber06,Kor09,Duc09}
{\em i.e.\ } spin polarization waves with specific 
wavevectors $(p_x,p_y) = (4 m \alpha, 0 )$.

Charge currents flowing through spin-orbit coupled
diffusive metals can generate finite spin accumulations.~\cite{Ede90,Lev85}
This magnetoelectric effect
achieves one of the main goals of spin-orbitronics -- creating a steady-state,
finite magnetization solely by applying an
external electric field. The direction
of polarization depends on the direction of the electric field and
on the spin-orbit interaction.
An electric field in $x$-direction leads to an
accumulation in $y$- or $x-$direction for linear Rashba [Eq.~(\ref{eq:rashbaH})]
or Dresselhaus [Eq.~(\ref{eq:dresselH})]
interaction, respectively. The magnetoelectric effect in
presence of both Rashba and Dresselhaus interaction has been investigated
in Refs.~\cite{Cha02,Tru07} which predicted that the CISP is
given by the uncorrelated sum of the two accumulations generated by
the Rashba
and Dresselhaus SOI independently of one another. In particular,
these predictions imply 
a finite accumulation at $\alpha = \pm \beta \ne 0$, whereas symmetry 
considerations (to be discussed below) require the vanishing of CISP
at this point. This motivates us to revisit this issue. 

The purpose of this paper is twofold. First, we revisit the Edelstein
magnetoelectric effect in presence of both Rashba and Dresselhaus linear
spin-orbit interaction. Contrarily to Refs.~\cite{Cha02,Tru07}, we find
that there is no CISP in any direction at $\alpha=\pm\beta$. However, the spin accumulation
of Refs.~\cite{Cha02,Tru07} is recovered at
an infinitesimally small distance 
away from the singular point $\alpha = \pm \beta$ in infinite systems.
Our second goal is therefore, and perhaps more importantly,
to figure out to what extent
phenomena occurring specifically at $\alpha = \pm \beta$ are physically
relevant. To that end, we consider three possible deviations from the
treatment of magnetoelectric effects given in Refs.~\cite{Ede90,Cha02,Tru07} in the form of
(i) finite-size effects, (ii)
the presence of a cubic Dresselhaus interaction
\begin{equation}\label{eq:dressel3H}
H_{\rm 3D} = \gamma(p_y p_x^2 \sigma_y - p_x p_y^2 \sigma_x) \,
\end{equation}
which is always there whenever a linear Dresselhaus interaction is present,
and (iii) an AC electric field.
We find that spin accumulations are suppressed
over parametric ranges given in each of these three instances by
(i) $|\alpha|-|\beta| \lesssim 1/mL$ which depends only
on the linear system size $L$, and not on the elastic mean free path $\ell$,
(ii)
$|\alpha|-|\beta| \lesssim \gamma p_{\rm F}^2$, and (iii)
$|\alpha|-|\beta| \lesssim \hbar \sqrt{\omega/m p_{\rm F}\ell}$.

There is a symmetry at $\alpha = \pm \beta$ that is responsible for
the vanishing of the magnetoelectric effect. In order to
expose that symmetry, we first
note that a linear SOI can be considered as a non-abelian SU(2) gauge
field with components
\begin{equation}
A_x= -m(\alpha \sigma_y+\beta \sigma_x),\quad A_y =m( \alpha \sigma_x +\beta \sigma_y) ,\quad A_z=0.
\label{EQ:SU_G}
\end{equation}
The corresponding field strength has only two nonzero 
components,
\begin{equation}
F_{xy}=-F_{yx}=i [A_x, A_y]=-m^2(\alpha^2-\beta^2)\sigma_z \, .
\end{equation}
They vanish for $\alpha=\pm \beta$.  Alternatively,
 one can consider the rotated Hamitonian given below in~Eq.(\ref{EQ:rotH})
 for $\alpha=\beta$ and perform the unitary transformation $U=e^{i\sigma_x \pi/2}$ to obtain
\begin{equation}
H=
\begin{pmatrix}
H_+ & 0 \\
0&H_-
\end{pmatrix},
\end{equation}
where $H_\pm=\frac{p^2}{2 m} \pm 2 \alpha p_x+V$. We thus see that the SU(2) gauge field reduces to two conventional U(1) gauge fields in the Hamiltonians $H_\pm$. This U(1) field is a pure gauge field, implying vanishing spin conductance.
To show this, one can for instance
consider the linear response expression for the spin conductance in a two-terminal mesoscopic sample~\cite{Baranger89}
\begin{equation}\label{eq:gmu}
G_\mu=\int_{C_i,C_j} {\rm d}{\bf x} {\rm d}{\bf x}'
{\rm Tr} [G^R({\bf x},{\bf x}') J_i' G^A({\bf x}',{\bf x}) J_j^\mu],
\end{equation}
where the trace is over the spin degree of freedom, the integrals are performed over
cross-sections $C_{i,j}$ of the two leads connecting the system to external terminals and
the current operators
$J_i=(i\nabla_i - {\bf A})/m$, $J_j^\mu=\{J_j,\sigma_\mu\}$. We write
$G^{R,A}({\bf x},{\bf x}')=g^{R,A}({\bf x},{\bf x}')e^{\pm i {\bf A}\cdot ({\bf x-x'})}$, where $g^{R(A)}$ is the retarded (advanced) Green's function of the system in the absence of SOI. For $\alpha=\pm \beta$, 
one can gauge the SOI
field out of the current operators via the transformation
\begin{equation}
e^{i {\bf A}\cdot ({\bf x-x'})} J_i e^{- i {\bf A}\cdot ({\bf x-x'})}=\frac{i}{m}\nabla_i \, ,
\end{equation}
which simultaneously gauges out the spin dependence of the Green's function in Eq.~(\ref{eq:gmu}).
We thus obtain ($\mu=x,y,z$)
\begin{eqnarray}
G_\mu=\int_{C_i,C_j} {\rm d}{\bf x} {\rm d}{\bf x}' \nabla_j' g^R({\bf x},{\bf x}') \nabla_i' g^A({\bf x}',{\bf x})  {\rm Tr}[\sigma_\mu]=0 \, .
\end{eqnarray}
It is straightforward to
see that this gauge argument also applies to CISP, since
the latter is given by a formula similar to Eq.~(\ref{eq:gmu}), with
the operator $J_j^\mu$ replaced by a Pauli matrix.

This article is organized as follows. In
Section~\ref{sec:electr-induc-spin}, we use spin- and charge coupled
diffusion equations to calculate the spin accumulation generated by a
charge current flowing in a bulk diffusive sample with Rashba and
Dresselhaus spin-orbit interactions. This approach allows us to
consider spin polarization in a finite size system
(Sec.~\ref{sec:electr-induc-spin-1}), an AC external electric field
(Sec.~\ref{sec:ac-solution}) and in the presence of a cubic Dresselhaus
interaction (Sec.~\ref{sec:cubic-dresselhaus}). Section
III presents numerical results on a tight-binding Hamiltonian confirming
our analytical predictions. A summary of our results and final comments
are given in the Conclusions section.

\section{Electrically induced spin polarization near $\alpha = \pm
  \beta$}\label{sec:electr-induc-spin}

We consider a disordered 2DEG with non-interacting electrons of mass
$m$ and charge $e$.  Choosing coordinates $\mathbf x, \mathbf y$ and
spin projections $\sigma_x$, $\sigma_y$ along the crystal axes $[1\bar 10]$
and $[110]$, respectively,\cite{caveat2} the system is described by the
Hamiltonian
\begin{align}
  \label{EQ:rotH}
H=\frac{p^{2}}{2m}+\boldsymbol{\Omega}(\mathbf{p})\cdot\boldsymbol{\sigma}+
  V({\bf x}).
\end{align}
Here, $\mathbf p = (p_x, p_y,0 )$ is the electron's momentum,
$\boldsymbol{\sigma} = (\sigma_x, \sigma_y, \sigma_z)$ is a vector of
Pauli matrices (we later use $\sigma^0 = \mathbb{1}$), and
\begin{equation}
\boldsymbol{\Omega}
(\mathbf p)= \sum_{k,j=1}^3\Omega_{kj} p_j \mathbf e_k = (- (\alpha -
\beta) p_{y}, (\alpha + \beta) p_{x}, 0)
\end{equation}
is the internal field due to
Rashba and linear Dresselhaus SOI given in 
Eqs.~(\ref{eq:rashbaH}) and~(\ref{eq:dresselH}), with strength
$\alpha $ and $\beta$, respectively. The disorder potential $V$ is
due to static, short-ranged and randomly distributed impurities
leading to a mean free path $\ell = p_F \tau /m $, where $\tau$ is the elastic
scattering time and $p_F$ the Fermi momentum. The interplay of
disorder and SOI is characterized by dimensionless parameters
$\xi_\alpha = 2 \alpha p_F \tau$,
$\xi_\beta = 2 \beta p_F \tau$ ($\hbar \equiv 1$)
measuring the spin precession angle due to Rashba and Dresselhaus SOI
between two consecutive scatterings at impurities.
Our treatment presupposes $\xi_{\alpha,\beta} \ll 1$, which ensures that spin
distribution functions vary slowly everywhere across the sample.

The coupled spin and charge excitations of the Rashba/Dresselhaus
spin-orbit coupled 2D electron gas obey the following integral
equation (summation over doubly--occurring indices is assumed)
\begin{eqnarray}
S_\mu ({\bf r},\omega)= \int \frac{{\rm d}{\bf r}'}{2 m \tau} {\rm Tr}
[ \sigma_\mu G^R_E ({\bf r},{\bf r}')\sigma_\nu G^A_{E-\omega}({\bf r}',{\bf r}) ] S_\nu ({\bf r}',\omega)
\nonumber ,
\end{eqnarray}
where $S_{x,y,z}({\bf r}, \omega)$ and $S_0({\bf r}, \omega)=n({\bf r},
\omega)$ are the spin and charge distribution functions, respectively. We obtain diffusion
equations in the presence of both Dresselhaus and Rashba SOI by
gradient expanding this integral equation.  For $\beta =0$,
these equations have been derived using
diagrammatic perturbation theory~\cite{Bur04}, kinetic equations~\cite{Mis04}
and quantum Boltzmann
equation approaches~\cite{Ada05}. For finite $\alpha$ and $\beta$ we
obtain the same diffusion equations as in Ref.~\onlinecite{Ber06}
which we rewrite here for convenience
\begin{subequations}\label{DE}
\begin{eqnarray}\label{DE0}
\partial_t n &=& D\nabla^2 n + K_{s-c}^x \partial_x S_y - K_{s-c}^y \partial_y S_x \, , \\
\label{DE1}
\partial_t S_{x}  &=&  D \nabla^2   S_{x} -   K_{s-c}^y \partial_y n
- K_p^x \partial_{x} S_z  - \Gamma_x S_{x} \, , \\
\label{DE2}
\partial_t S_{y}  &=& D \nabla^2  S_{y} +  K_{s-c}^x  \partial_{x} n
- K_p^y \partial_{y} S_z  - \Gamma_y  S_{y} \, , \qquad \\
\label{DE3}
\partial_t S_z &=&  D \nabla^2 S_z +   K_p^y\partial_{y} S_{y} +  K_p^x
\partial_{x} S_{x} - \Gamma_z  S_z.
\end{eqnarray}
\end{subequations}
Here the spin-charge couplings $K_{s-c}^{x,y}$, precession couplings
$K_{p}^{x,y}$ and spin relaxation rates  $\Gamma_{x,y}$ are given by
\begin{subequations}\label{couplings}
\begin{eqnarray}
K_{s-c}^x&=& 4m^2 D \tau (\alpha - \beta)^2(\alpha + \beta)  \, ,
\\
K_{s-c}^y&=& 4m^2 D \tau (\alpha + \beta)^2(\alpha - \beta) \,  ,  \\
K_p^x&=& 4m D (\alpha + \beta) \,  ,  \\
 K_p^y&=&  4m D(\alpha - \beta) \, ,  \\
\Gamma_x &=& 1/\tau_x =  4m^2 D (\alpha+ \beta)^2    ,  \\
\Gamma_y &=& 1/\tau_y = 4m^2 D (\alpha- \beta)^2  ,\\
\Gamma_z &=& \Gamma_x +\Gamma_y \,  .
\end{eqnarray}
\end{subequations}

For a homogeneous sample with a homogeneous charge current density, it
is tempting to assume homogeneous spin accumulations and
ignore all partial derivatives of $S_\mu$ to obtain
\begin{subequations}
\begin{eqnarray}
\label{eq:S-infinite-a}
S_x & \propto& -K_{s-c}^y \tau_x \partial_y n =-(\alpha-\beta)
\tau \partial_y n\, , \\
\label{eq:S-infinite-b}
S_y & \propto & K_{s-c}^x \tau_y \partial_x n = (\alpha+\beta)
\tau \partial_x n\, , \\
\label{eq:S-infinite-c}
S_z & = & 0 \, .
\end{eqnarray}
\end{subequations}
for the bulk Edelstein CISP. We note the cancellation of the
potentially singular $(\alpha \pm \beta)$ factors in $\tau_{x,y}$.  For
$\alpha\rightarrow\pm \beta$ the spin-charge couplings go to zero but
this behavior seems to be cancelled by the diverging spin relaxation
time to give finite spin accumulations at $\alpha=\pm \beta$.
However the same approach for $\alpha$ set to $\beta$ from the outset
produces vanishing spin accumulations. The main reason behind this
inconsistency is that one spin
relaxation time of the system diverges as $\alpha\rightarrow\pm \beta$.
However, in a real, finite-sized system,
the spin relaxation time is bounded from above by the typical time
to escape to the leads. This is so, because leads provide spin (and charge)
relaxation, which for $\alpha = \pm \beta$ becomes the dominant
spin relaxation mechanism. Finite-sized effects are thus expected to
induce a smooth crossover to zero CISP as $\alpha \rightarrow \pm \beta$. 
In the next section, we show that this is indeed the case. 

\subsection{Electrically induced spin polarization in finite systems}\label{sec:electr-induc-spin-1}

We assume a rectangular sample with SOI
attached to two external reservoirs defining the current direction,
and bounded by vacuum otherwise. We obtain for the charge distribution function
\begin{equation}
\label{eq:charge-dens}
n(E)=2(1-x/L){\mathcal F}(E-eV) +(2x/L) {\mathcal F}(E),
\end{equation}
where ${\mathcal F}(E)$ is the Fermi function.  The appropriate
boundary conditions are that the spin accumulations vanish at the
reservoirs and the normal component of the spin current vanishes at
the hard wall boundaries~\cite{Ada07,Tse07}.

Solving the diffusion equations we obtain the maximum spin
accumulation within the SOI region for an electric field along the
$x$-direction:
\begin{subequations}
\label{s_an_finsize}
\begin{eqnarray}
S_y&=&S_\mathrm{2DEG}\big(1-1/\cosh(mL\vert \alpha -\beta\vert/\hbar) \big) \, , \\
S_\mathrm{2DEG}&=&(\alpha +\beta )\tau \frac{\mathrm{d}n}{\mathrm{d}x} \, .
\end{eqnarray}
\end{subequations}
For a field in the $y$-direction, one has the same behavior
for $S_x$ instead of $S_y$, with$|\alpha+\beta|$ in the argument of the
cosh and $S_\mathrm{2DEG} = -(\alpha-\beta)\tau{\rm d}n/{\rm d}y$. Eq.~(\ref{eq:charge-dens}) shows that
the Edelstein CISP goes smoothly to zero for $\alpha=\pm \beta$,
with the width of the crossover set solely by the system
size, generating a singular behavior only as $L \rightarrow \infty$. The
size of the crossover region is in particular independent
of the mean free path $\ell$, hence of the strength of the
impurity potential, since in our regime, $\xi_{\alpha,\beta} \ll 1$, the
spin-orbit relaxation length is independent of disorder. Away from
$\alpha = \pm \beta$, one recovers the standard CISP $S_\mathrm{2DEG}$ predicted
in Refs.~\cite{Cha02,Tru07}.
The validity of Eq. (\ref{eq:charge-dens}) is illustrated numerically
below in Fig.~\ref{fig:A=B_nums}.

\subsection{AC-field induced spin polarization}\label{sec:ac-solution}

We next discuss the frequency dependence of CISP due to an AC electric field
within the framework of the diffusion equations (\ref{DE0}).
For $\alpha=\pm\beta$, this problem has already been addressed by Raichev~\cite{Rai07},
and we revisit it briefly only for completeness.
In an infinite system the polarization is
spatially homogeneous such that all derivatives of $S_\mu$ in
Eqs.~(\ref{DE1})-(\ref{DE3}) vanish. The resulting bulk polarization
then satisfies
\begin{subequations}
  \label{eq:DE-matrix}
\begin{eqnarray}
(-i\omega + \Gamma_x) S_x &=& -K^y_{s-c} \partial_y n \, , \\
(-i\omega + \Gamma_y) S_x &=& K^x_{s-c} \partial_x n \, , \\
(-i\omega + \Gamma_z) S_z &=& 0 \, .
\end{eqnarray}
\end{subequations}
Further neglecting the influence of SOI on $n$
one finds from Eq.~(\ref{eq:charge-dens}) that $\nabla n = -2 \nu e
\mathbf E$ and thus
\begin{subequations}
\label{eq:ac-pol}
\begin{eqnarray}
S_x &=&  2 \nu e E_y (\alpha-\beta) {\rm Re} [\Gamma_x/(\Gamma_x -i \omega)]   \, , \\
S_y &=& -2 \nu e E_x (\alpha+\beta) {\rm Re} [\Gamma_y/(\Gamma_y -i \omega)] \, , \\
S_z &=& 0
\end{eqnarray}
\end{subequations}
for finite but small $\omega \tau \ll 1$. This result has been found
in \cite{Mal05, Duc07}. As for finite-sized systems, we see that both spin accumulations
vanish at $\alpha = \pm \beta$ and that the result of Refs.~\cite{Cha02,Tru07} is only
recovered at $|\alpha|-|\beta| \sim \sqrt{\omega/m p_{\rm F} \ell}$. In the limit $\omega \tau \rightarrow 0$
the polarization vanishes at the singular points only.

\subsection{CISP in presence of a cubic Dresselhaus interaction}\label{sec:cubic-dresselhaus}

A linear Dresselhaus SOI, Eq.~(\ref{eq:rashbaH}), is always
accompanied by a cubic Dresselhaus interaction,
Eq.~(\ref{eq:dressel3H}), whose strength might or might not be much
weaker than that of the linear SOI.  Because the presence of a cubic
Dresselhaus SOI breaks U(1) symmetry at $\alpha=\pm\beta$, whose
presence is crucial to the vanishing of the CISP, we investigate in
this paragraph the effect that a cubic Dresselhaus SOI has on the CISP
close to those points.

\begin{figure}
  \centering
  \includegraphics[width = 0.95 \linewidth]{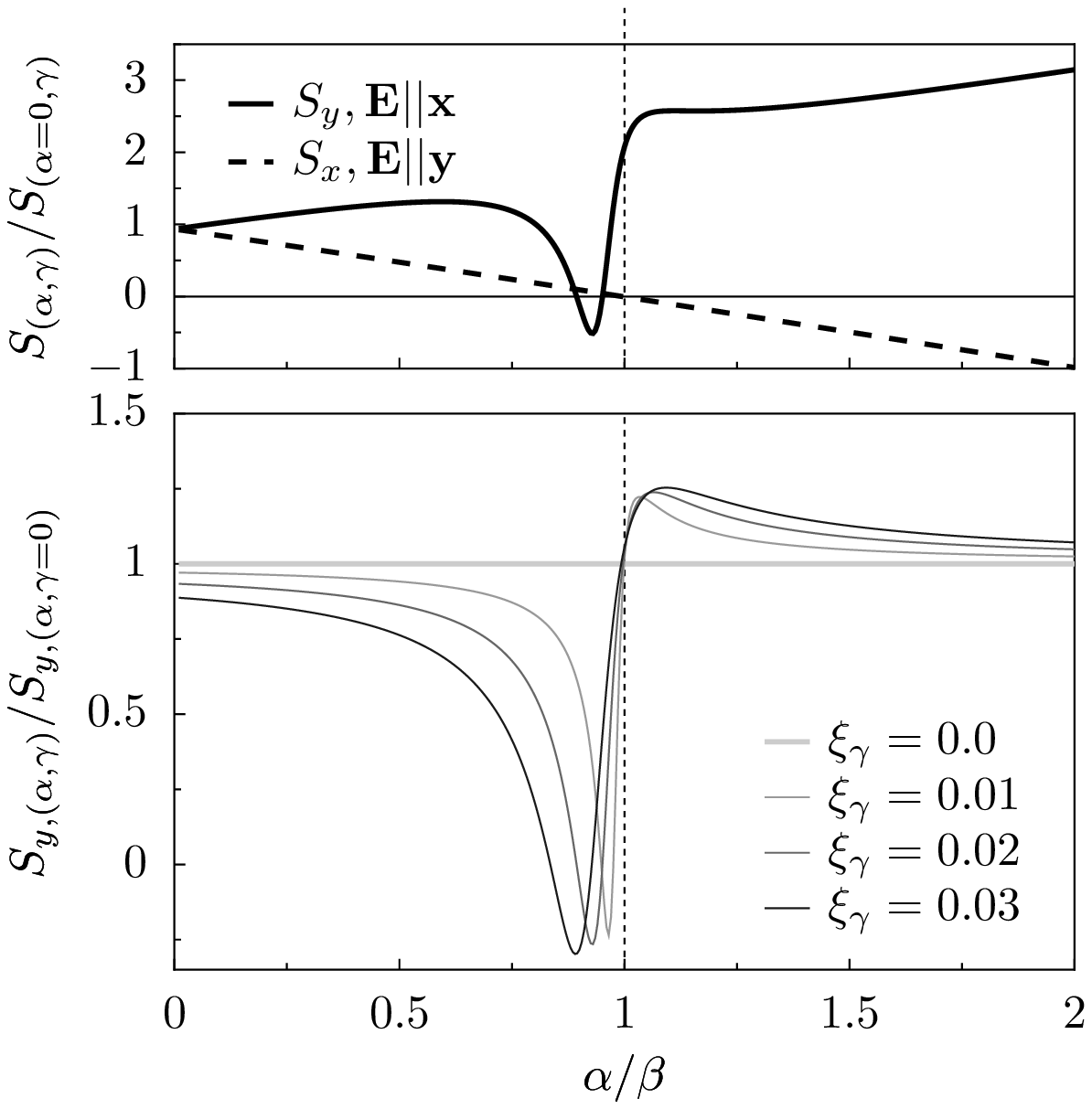}
  \caption{Upper panel: Spin polarization $S_{x,(\alpha)} / S_{x,(
      \alpha=0)}$ for $\mathbf E || [110]$ (dashed) and
    $S_{y,(\alpha)} / S_{y,( \alpha=0)}$ (solid line) for $\mathbf E
    || [1\bar 10]$ as a function of Rashba SOI $\alpha/ \beta$ for
    $\xi_{\beta}= 2 \beta p_F \tau=0.1$ and $\xi_\gamma = 0.02$.
    Lower panel: $\alpha$-dependence of the normalized spin
    polarization $S_{y,(\alpha, \gamma)} / S_{y, (\alpha, \gamma=0)}$
    for $\mathbf E || [110]$, $\xi_{\beta}= 2 \beta p_F \tau=0.1$,
    and $\xi_\gamma = 0.0, 0.01,0.02, 0.03$.}
  \label{spin-pol-cubic}
\end{figure}

If the cubic contributions are weak we still expect a suppression of
the CISP at $\alpha=\pm \beta$ and that the additional spin relaxation
due to $H_{3D}$ renders the point $\alpha = \pm \beta$ non-singular in
the absence of boundary effects and at zero frequency.  In the
coordinates chosen in Eq.~(\ref{EQ:rotH}) the cubic term in the SOI
Hamiltonian is
\begin{align}
  H_{3\text{D}} & =\frac{1}{2} \gamma \left(p_y^2 - p_x^2\right) 
  \left(p_x \sigma _y-p_y \sigma _x\right). 
\end{align}
which has to be incorporated into the diffusion
Eqs.~(\ref{DE1})-(\ref{DE3}). The relevant relaxation rates
$\Gamma_\mu$ and spin-charge couplings $K_{s-c}^\mu$ have been
calculated in Ref.~\onlinecite{Ave02} and Ref.~\onlinecite{Mal05},
respectively. In our notation they are given by
\begin{subequations}
\begin{align}
  \label{eq:DE-coefficients-cubic}
\Gamma_{x,y} &= \frac{\left(\xi_{\alpha}\pm
    \xi_{\beta}\right)^{2}}{2\tau} - \frac{\xi_{\gamma} \left(\xi_{\beta}\pm
   \xi_{\alpha}\right)}{4\tau}+\frac{\xi_{\gamma}^{2}}{16\tau} \\ \notag
&= Dm^{2} \left[4(\alpha\pm\beta)^{2} - 2 (\beta \pm \alpha)\gamma
p_{F}^{2}+\frac{1}{2} \gamma^{2}p_{F}^{4} \right] \\
  \label{eq:DE-coefficients-cubic2}
K_{s-c}^{x,y} &=
(\alpha\pm\beta)
\frac{(\xi_\alpha\mp\xi_\beta)^{2}}{2} \pm \frac{3}{4}(\alpha^{2}-\beta^{2}) \tau
p_{F} \xi_{\gamma} \notag \\ &+ \frac{1}{16} ( 3
\alpha \mp \beta)\xi_{\gamma}^{2} \pm \frac{3\xi_{\gamma}^{3}}{256\tau
  p_{F}} ,
\end{align}
\end{subequations}
where $\xi_\gamma = 2 \gamma p_F^3 \tau$ and the upper (lower) sign
applies to the $x$ ($y$) component.  In the presence of cubic SOI
the relation $K_{s-c}^{x,y} = \tau \Gamma_{y,x} (\alpha \mp
\beta)$, which led to the cancellation of divergent terms in
  Eq.~(\ref{eq:S-infinite-a}-\ref{eq:S-infinite-c}), no longer
holds. The polarization is 
given by Eq.~(\ref{eq:DE-matrix}),
\begin{align}
\label{eq:pol-cubic}
S_\mu = 2 \nu e \, \epsilon_{z \mu \nu} \, \Gamma_\mu^{-1} \, K^\nu_{s-c}
\, E_\nu \, ,
\end{align}
where now $\Gamma_{x,y}$ and $K_{s-c}^{x,y}$ are given in
Eq~(\ref{eq:DE-coefficients-cubic2}), and $\epsilon_{z \mu \nu}$ is
the totally antisymmetric tensor of order three.  The CISP is a
rational function of $\xi_{\alpha,\beta,\gamma}$.
Fig.~\ref{spin-pol-cubic} shows the behavior of $S_{x,y}$ in the
presence of weak cubic Dresselhaus SOI ($\xi_\gamma = 2 \gamma p_{\rm
  F}^3 \tau \ll \xi_\beta$), as a function of $\alpha/\beta$. In this
case, the polarization $S_y$ does not vanish precisely at
$\alpha=\beta$ but shows a feature in the vicinity 
of this point. The minimum and maximum around the feature are at
$\alpha = \beta (1 \mp \xi_\gamma/(\xi_\beta 2 \sqrt{2}))$.  The zeros
are at $\alpha = \beta (1 - \xi_\gamma/2 \xi_\beta)$ and $\alpha
= \beta (1 - \xi_\gamma/4 \xi_\beta)$. Thus we conclude that a
weak cubic Dresselhaus interaction regularizes the singularity of the
CISP around $\alpha = \pm \beta$. The suppression of the CISP occurs
over a width $\propto \gamma p_{\rm F}^2$ around $\alpha = \pm
\beta$. The predicted analytical dependences of $S_\mu$ on Rashba and
Dresselhaus SOI strengths in Eqs. (\ref{eq:DE-coefficients-cubic}),
(\ref{eq:DE-coefficients-cubic2}) and (\ref{eq:pol-cubic}) may serve
as guidance when attempting to tune quantum wells to the symmetry
points $\alpha=\pm \beta$ and demonstrate the vanishing of the CISP
due to linear SOI at this point.

\section{Numerical simulations}\label{sec:numerics}

We now perform quantum tranport simulations
demonstrating the suppression of the CISP
around the singular point $\alpha =\pm\beta$ for finite size
geometries. To this end
we consider coherent electron transport in a disordered quantum wire of width $W$
with linear Rashba and Dresselhaus SOI. For the calculations we use a tight-binding
version of the Hamiltonian~(\ref{EQ:rotH}) that we obtain from a discretization of the system
on a square grid with lattice spacing $a$. The Hamiltonian
is $H=H_0+H_{\rm so}$ with
\begin{subequations}
\begin{eqnarray}
\label{num-hamiltonian1}
H_0 &=& -t\sum_{q,\sigma}(c^\dagger_{q,\sigma} c_{q+{\hat x},\sigma}+c^\dagger_{q,\sigma}
c_{q+{\hat y},\sigma}+h.c.)   \\
&&+\sum_{q,\sigma}U_q c^\dagger_{q,\sigma} c_{q,\sigma}, \nonumber \\
H_{\rm so} &=& \sum_q
[-(t_{\rm R}+t_{\rm D})(c^\dagger_{q,\uparrow} c_{q+{\hat x},\downarrow}
-c^\dagger_{q,\downarrow} c_{q+{\hat x},\uparrow})  \\
\label{num-hamiltonian2}
&& + \mathrm{i}(t_{\rm R}-t_{\rm D})( c^\dagger_{q,\uparrow} c_{q+{\hat y},\downarrow}
+c^\dagger_{q,\downarrow} c_{q+{\hat y},\uparrow})+h.c.] \, . \nonumber
\end{eqnarray}
\end{subequations}
Here $c^\dag_{q,\sigma}$ ($c_{q,\sigma}$) creates (annihilates) an electron
with spin $\sigma=\uparrow$ or $ \downarrow$ in $\hat z$-direction on site $q=(q_x,q_y)$.
The vectors ${\hat x}$ and ${\hat y}$ have 
length $a$ and point in $x$ and $y$ directions, 
$t=1/2ma^2$ denotes the hopping energy, while 
$t_{\rm R} = \alpha/2a$ and $t_{\rm D} = \beta/2a$
are the Rashba and Dresselhaus SOI strength, respectively, in terms of which
the spin-orbit lengths are
given by $\ell_{\rm so}^\mathrm{R/D}=\pi a t/t_{\rm R,D}$.
We furthermore include spin-independent disorder of Anderson type in the 
region of length $L$,
where the on-site energies are randomly box-distributed with 
$U_q \in [-U/2,U/2]$. The disorder strength $U$ determines
the elastic mean free path $\ell \approx 48 a t^{3/2}\sqrt{E_\mathrm{F}}/U^2$, 
which we tuned to values large enough that the system is not localized,
but much smaller than the size of the
disordered region in all our simulations

We obtain the local electron and spin densities
\begin{subequations}
\begin{eqnarray}
n=-\mathrm{i}\mathrm{Tr}\left[G^<(q,q)\right] \, , \\
S_\mu=-\mathrm{i}\mathrm{Tr}\left[\sigma_\mu G^<(q,q)\right]
\end{eqnarray}
\end{subequations}
at site $q$ by numerically computing the lesser Green function $G^{<}(q,q)$.
To this end we employ an efficient recursive lattice Green function method 
based
on matrix-reordering algorithms as described in Ref.~\onlinecite{Wim08}.
We calculate averaged quantities $\langle S_i\rangle$ and $\langle n\rangle$,
over several thousands of disorder
configurations and over a rectangular region in the center of
the disordered part of the wire. 
\begin{figure}[tb]
    \includegraphics[width=0.93 \columnwidth]{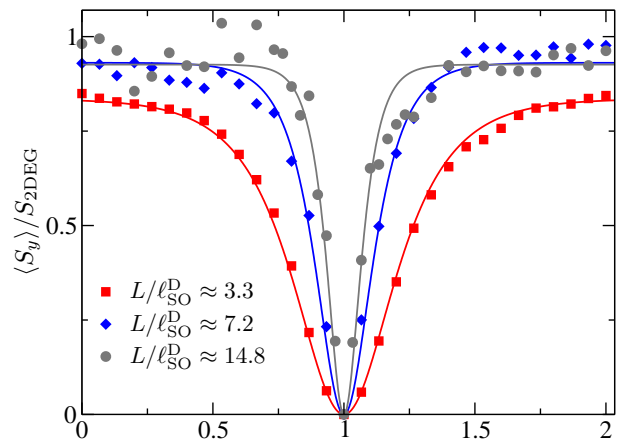}
   \caption{(color online) Normalized spin accumulation 
$S_y /S_\mathrm{2DEG}$ as a function of $\alpha /\beta$ for fixed 
$\beta / 2a = t_\mathrm{D}=0.15t$ (giving 
$\ell_{\rm so}^\mathrm{D}\approx 21a$), $U = 2t$ (giving 
$\ell\approx 8.5a$) and Fermi energy $E_\mathrm{F} = 0.5t$,
for different 
linear system size $W=L=70 a$ (red squares), $150a$ (blue diamonds) and
$310a$ (grey circles). Data are averaged over $5000$ disorder configurations. 
The solid lines are the theoretical prediction,
Eq.~(\ref{s_an_finsize}), with renormalized 
bulk spin accumulation and system size,
$S_\mathrm{2DEG}   \rightarrow \delta_\mathrm{fit}S_\mathrm{2DEG}$
and $L \rightarrow L_\mathrm{fit}$ with
$\delta_\mathrm{fit}\approx 0.84$, $L_\mathrm{fit}\approx 39.3a$
for $L=70a$, $\delta_\mathrm{fit}\approx 0.93$, 
$L_\mathrm{fit}\approx 69.7a$ for $L=150a$ and 
$\delta_\mathrm{fit}\approx 0.93$, $L_\mathrm{fit}\approx 117.1a$ 
for $L=310a$. The electric current is in 
the direction $\hat{x}\parallel [ 1\bar{1}0]$.
\label{fig:A=B_nums} }
\end{figure}
We compare numerical data
with the analytical prediction of Eq.~(\ref{s_an_finsize}).
In Fig.~\ref{fig:A=B_nums} we show the normalized, spatially
averaged spin accumulation, 
$\langle S_y\rangle/S_{\rm 2DEG}$,
as a function of $\alpha /\beta$ varying the linear system
size $L$.
As expected, we find complete suppression of 
$\langle S_y \rangle$ at $\alpha =\beta$,
in agreement with Eq.~(\ref{s_an_finsize}).
Moreover, 
the pronounced  dip around $\alpha =\beta$ becomes sharper and sharper
as $L$ increases, and the numerical data are in good qualitative agreement
with the predicted line shape, Eq.~(\ref{s_an_finsize}),
in particular, they have the same parametric dependence. 
The agreement becomes even quantitative if one normalizes 
the system size and the bulk spin accumulation in Eq.~(\ref{s_an_finsize}), 
as is done in Fig.~\ref{fig:A=B_nums}. We justify this normalization
by the effective reduction of the spin-orbit interaction 
in confined systems with homogeneous SOI,~\cite{Ale01}
and the fact that
$\ell_{\rm so}^\mathrm{D} \approx 2.5 \ell$ is barely in the regime
of validity $\ell_{\rm so} \gg \ell$ of Eq.~(\ref{s_an_finsize}).
This leads
to smaller bulk spin accumulations and a longer spin relaxation length 
$L_{\rm s} = \sqrt{D \tau_{x,y}}$ than 
the case in which the conditions $\xi_{\alpha,\beta} \ll 1$ and 
$L_{\rm s}\ll L$ 
are completely fulfilled, and qualitatively
explains the renormalization of the effective 
system length and the bulk spin accumulation. 
We also note that finite-sized effects lead to deviations 
from our estimates $\ell \approx 48 a t^{3/2} \sqrt{E_{\rm F}}/U^2$ 
for the elastic mean free path, and that numerical estimates based on
the average inverse participation ratio~\cite{Pri98} 
of eigenstates systematically give a larger value for $\ell$ for which
$\ell/\ell_{\rm so}^{\rm D}\simeq 0.55$.
\begin{figure}[tb]
    \includegraphics[width=0.95 \columnwidth]{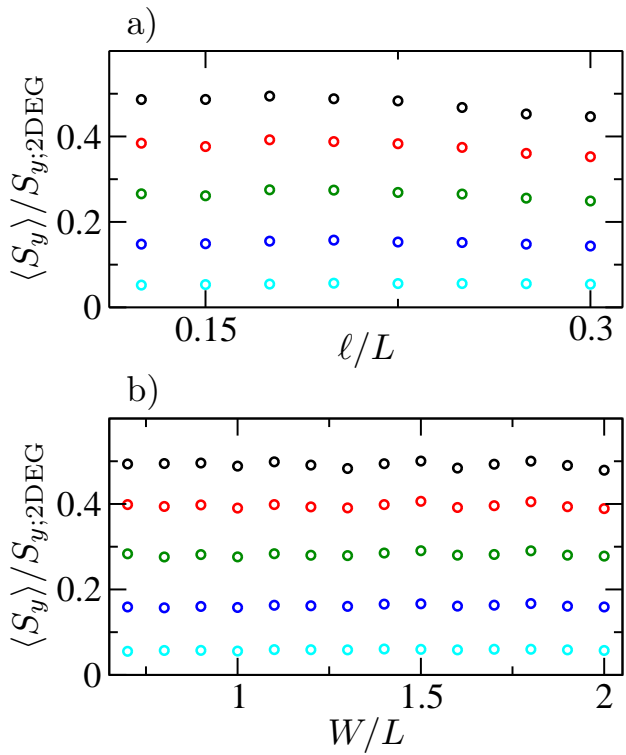}
  \caption{(color online) Disorder-averaged normalized spin accumulation
  $\langle S_{y}\rangle /S_{y;\rm 2DEG}$, with $S_{y;\rm 2DEG}=\alpha \tau (\mathrm{d}n/\mathrm{d}x )$,
  as a function of a) the mean free
  path $\ell$ (for fixed width $W=50a$) and b) the width $W$ of the wire
(for fixed $U = 2t$, $\ell \approx 8.5a$). The electric current is in 
the direction $\hat{x}\parallel [ 100]$.
Different data sets corresponds to different values of $\beta /\alpha=n/15$, 
$n=10$ (black circles), 11 (red), 12 (green),13 (dark blue) and 
$14$ (light blue). In both panels, other parameters are fixed at
   $t_\mathrm{R}=0.15t$, $E_F = 0.5t$,
   $L = 40 a$ and data have been averaged over 3000 disorder configurations.
\label{fig:A=B_nums2} }
\end{figure}

According to  Eq.~(\ref{s_an_finsize}), the suppression of the CISP
is independent of the strength of disorder / 
the elastic mean free path of the sample, as long as one stays 
in the diffusive regime.
This prediction is supported by our numerical calculations. 
We find that the spin accumulation stays approximately constant with
respect to the electronic mean free path. This is shown in
Fig.~\ref{fig:A=B_nums2}a.
In Fig.~\ref{fig:A=B_nums2}b we moreover confirm that the CISP is independent
of the width $W$ of the rectangular SOI region for $W\ge L$.
However, we expect a width dependence in the form of a
reduction of the
CISP upon reducing $W$, when D'yakonov-Perel' spin relaxation~\cite{Dya71b} 
begins
to be reduced and finally suppressed due to the lateral 
confinement.~\cite{Schaepers06,Scheid09}

\section{Conclusions}

In this work we have studied the electrically induced and spin-orbit mediated
spin accumulation in two-dimensional diffusive conductors with emphasis on finite-size
and finite-frequency effects. In the thermodynamic limit of extended systems
with (linear) Rashba and Dresselhaus SOI the Edelstein magnetoelectric effect gives rise to
finite spin accumulation up to suppression at the singular point $|\alpha| = |\beta|$.
However, in many experimentally relevant systems, additional time, respectively, energy scales
come into play, such as in tranport (i)  through mesoscopic samples of finite size,
(ii) in the AC regime and (iii) through samples with cubic Dresselhaus SOI.
We have shown, both analytically and numerically, that in these situations the
singularity in the  spin accumulation at $|\alpha| = |\beta|$ is widened to 
a dip.
This suppression of the spin accumulation over a finite $\alpha/\beta$-range
close to $\alpha=\pm \beta$ may have interesting implications with regard 
to other
phenonema based on the Dyakonov-Perel spin relaxation mechanism.
As but one consequence, finite-size effects may render the
spin-field-effect transistor proposed in Ref.~\onlinecite{Sch03} for
$|\alpha| = |\beta|$ effectively operative even if the two
linear SOI are not precisely equal. This is so, because the spin
rotation along two different trajectories with the same
endpoints remains the same, even away from $|\alpha| = |\beta|$,
if the trajectories are not too long. This is reflected in
the finite width $|\alpha| - |\beta| \lesssim 1/mL$ of the
CISP lineshape given in Eq.~(\ref{s_an_finsize}).
Furthermore, given that spin helices also emerge from Eqs.(\ref{DE0}) and
(\ref{couplings})~\cite{Ber06,Kor09}, we conjecture 
that it is either finite-size
effects or the presence of a cubic Dresselhaus SOI, or both,
that render persistent spin helices excitable some distance away
from $\alpha=\pm \beta$, and thus experimentally observable.

While the present analysis is based on diffusive charge carrier motion, it
would be interesting to investigate ballistic
mesoscopic systems and see
whether our results apply there
or if our analysis has to be extended. Work along these lines is in progress.

\section*{ACKNOWLEDGMENTS}

We thank John Schliemann for a careful reading of the manuscript.
PJ thanks 
the physics department of the Universities of Basel and
Geneva for their hospitality at various stages of this project and
acknowledges the support of the National Science Foundation under
    Grant No. DMR-0706319. DL and MD acknowledge financial support
from the Swiss NF and the NCCR Nanoscience Basel. IA 
is supported by the funds of the Erdal In\"{o}n\"{u} Chair of Sabanci 
University.
IA and KR thank the
Deutsche Forschungsgemeinschaft for support within the cooperative
research center SFB 689, and
MS acknowledges support from the \emph{Studienstiftung des Deutschen Volkes}.
IA and PJ express their gratitude to the Aspen Center for Physics for its
hospitality.


\begin{thebibliography}{99}
\bibitem{caveat} We note that these equations also allow to compute the spin current due to 
magneto-electric  effects via  Fick's law for the spin current~\cite{Ada05},
$
J_s=-\nu D \nabla S \, .
$
One straightforwardly obtains $J_s\propto(\alpha p_F \tau)^2$.
\bibitem{caveat2} The transformation
is $p\rightarrow p'$ and $\sigma \rightarrow
  \sigma '$ with $p_x'=(p_x+p_y)/\sqrt{2}$, $p_y'=-(p_x-p_y)/\sqrt{2}$,
  $\sigma_x'=(\sigma_x+\sigma_y)/\sqrt{2}$ and
  $\sigma_y'=-(\sigma_x-\sigma_y)/\sqrt{2}$.
\end{thebibliography}
\end{document}